\documentstyle{elsart}

\begin{document}
\begin{frontmatter}
\title{TeV observations of Centaurus~A}
\author[ICRR]{G.P.~Rowell},
\author[Adelaide]{S.A.~Dazeley},
\author[ISAS]{P.G.~Edwards},
\author[Yamagata]{S.~Gunji},
%\author[TIT]{S.~Hara},
\author[YGU]{T.~Hara},
\author[ICRR]{J.~Holder},
%\author[Tokai]{J.~Jimbo},
\author[ICRR]{A.~Kawachi},
\author[ICRR]{T.~Kifune},
%\author[TIT]{H.~Kubo},
%\author[TIT]{J.~Kushida},
%\author[ICRR]{S.~Le~Bohec},
\author[Nagoya]{Y.~Matsubara},
\author[NAOJ]{Y.~Mizumoto},
\author[ICRR]{M.~Mori},
%\author[TIT]{M.~Moriya},
\author[Ibaraki]{H.~Muraishi},
\author[Nagoya]{Y.~Muraki},
\author[NAOJ]{T.~Naito},
\author[Tokai]{K.~Nishijima},
\author[TIT]{S. Ogio}
\author[Adelaide]{J.R.~Patterson},
\author[ICRR]{M.D.~Roberts},
\author[Nagoya]{T.~Sako},
\author[TIT]{K.~Sakurazawa},
\author[RIKEN]{R.~Susukita},
\author[Kanagawa]{T.~Tamura},
\author[TIT]{T.~Tanimori},
\author[Adelaide]{G.J.~Thornton}
\author[Ibaraki]{S.~Yanagita},
\author[Ibaraki]{T.~Yoshida},
\author[ICRR]{T.~Yoshikoshi},
%\author[Nagoya]{A.~Yuki}
\address[ICRR]{Institute for Cosmic Ray Research, University of Tokyo,
Tokyo~188-8502, Japan}
\address[Adelaide]{Dept. of Physics and Math. Physics,
University of Adelaide~5005, Australia}
\address[ISAS]{Institute of Space and Astronautical Science,
Kanagawa 229-8510, Japan}
\address[Yamagata]{Dept. of Physics, Yamagata University,
Yamagata~990-8560, Japan}
\address[TIT]{Dept. of Physics, Tokyo Institute of Technology,
Tokyo~152-8551, Japan}
\address[YGU]{Faculty of Management Information, Yamanashi Gakuin University,
Yamanashi~400-8575, Japan}
\address[Tokai]{Dept. of Physics, Tokai University,
Kanagawa~259-1292, Japan}
\address[Nagoya]{Solar-Terrestrial Environment Lab., Nagoya University,
Aichi~464-8601, Japan}
\address[NAOJ]{National Astronomical Observatory of Japan,
Tokyo~181-8588, Japan}
\address[Ibaraki]{Faculty of Science, Ibaraki University,
Ibaraki 310-8512, Japan}
\address[RIKEN]{Institute of Physical and Chemical Research,
Saitama 351-0198, Japan}
\address[Kanagawa]{Faculty of Engineering, Kanagawa University,
Kanagawa~221-8686, Japan}

%\thanks[list]{For a collaboration member list see {\tt http://icrhp9.icrr.u-tokyo.ac.jp}}
%\address{Institute for Cosmic Ray Research, University of Tokyo, Tokyo 188-8502}

\begin{abstract}
 We have searched for TeV gamma-rays from Centaurus~A and surrounding region out to $\pm 1.0^\circ$
 using the CANGAROO 3.8m telescope. No evidence for TeV gamma-ray emission was observed from the
 search region, which includes a number of interesting features located away from the tracking
 centre of our data. The 3$\sigma$ upper limit to the flux of gamma-rays above 1.5 TeV from an
 extended source of radius 14$^\prime$ centred on Centaurus A is $1.28\times 10^{-11}$ 
 photons cm$^{-2}$ s$^{-1}$.
\end{abstract}
\end{frontmatter}

\newcommand{\C}{\v{C}erenkov }

\section{Introduction}
 Centaurus~A (NGC 5128) is the closest known radio galaxy ($\sim$3.5 Mpc), and is
 considered a misaligned AGN with a jet orientation angle of $\sim 70^\circ$
 \cite{Tingay:1}. Variability in X-ray and low energy gamma-ray flux of up to an order 
 of magnitude on time scales of days to years has been observed \cite{Bond:1}. 
 The EGRET source 2EGJ1326-43, is considered to be 
 associated with Centaurus~A \cite{Steinle:1,Sreekumar:1}. 
 A number of upper limits and marginal claims for detection at TeV and PeV energies have also been reported.
 See \cite{Allen:1} and references therein.
% \cite{Grindlay:1,Carraminana:1,Allen:1,Allen:1,Bird:1}.

% In the present
% work, we have subjected these data to a search for point-like and extended TeV emission over a $\pm1^\circ$
% region which includes a large fraction of the EGRET 95\% error circle and the peak radio emission
% from the Northern Middle Lobe
% (NML). The NML is thought to be an end-point of the AGN jet interacting with the surrounding medium
%  \cite{Romero:1}.

 We used the CANGAROO 3.8m telescope \cite{Hara:1} in observations of Centaurus~A taken from 
 March to April 1995.  
 A total of 45 hours of ON and OFF source data were considered for analysis. 
 An earlier analysis \cite{Susukita:1} assumed a single point source at the tracking centre.
 However, the spatial extent and number of interesting features of the Centaurus~A region warranted an 
 extended source analysis, out to $\pm1^\circ$.

\section{Analysis}
 We have based our analysis on the method of \cite{Buckley:1} 
 in which 
 location cuts, recalculated at every grid position of the search, are combined with shape cuts to
 form a skymap of the ON--OFF significance.
 We find that due to camera edge effects, some adjustment of image cuts as a function of
 source location is necessary to optimise the cosmic-ray background rejection over the search
 region. Location cuts used are {\em asymmetry} and the normalised distance between the
 assumed and calculated source position of the \C image. Shape cuts are the image {\em width} and {\em length}.
 This method will be described in more detail in a later paper. The actual values of each cut were
 selected {\em a priori} using Monte Carlo simulations. We found that the total cut combination 
 provides a gamma ray acceptance of $\sim 40$\% and cosmic ray acceptance of $\sim 1$\% for point sources
 within $\pm 1^\circ$ of the camera centre.

 Three sites were considered as potential gamma-ray sources within the search: the tracking centre of these
 data based on the radio VLBI core position \cite{Johnston:1}, the unidentified EGRET source, and the 
 Northern Middle Lobe (NML) \cite{Romero:1}.
% About 86\% of the 95\% EGRET error circle was covered by our search. 
 The tracking centre and EGRET source
 were considered as both point-like and extended sources while the NML was considered as a point-like source
 only as it is close to the search boundary. For an extended source, the ON and OFF source
 counts were obtained by summing the events passing cuts using a suitably high number of assumed source positions 
 within the region of interest, taking care not to count an event more than once.  

% \begin{figure}
% \vspace{11cm}
% \special{psfile=../cenAmap.ps hscale=60 vscale=60 voffset=-350 hoffset=-290}
% \begin{center}
%   Figure 1. Skymap of ON--OFF excess for the Centaurus~A 1995 dataset. 
%   The coordinate origin (crossed circle) represents the tracking centre
%   (RA: 13$^h 25^m 29^s$, Dec: -43$^\circ 01^m 12^s$ (J2000)).
% \end{center}
% \end{figure}

 \begin{table}
  \begin{center}
  \caption{Summary of ON--OFF excesses and 3$\sigma$ upper limits for the 
    Centaurus~A region. Where appropriate, the source/search radius is indicated.}
  \label{tab:upplimits}
  \begin{tabular}{lcc} \hline \hline
   Feature                               & ON--OFF($\sigma$) & Flux($\geq 1.5$ TeV) ph cm$^{-2}$ s$^{-1}$ \\ \hline
   Tracking$^a$ (point)                  & +1.6   & $<5.45\times 10^{-12}$\\
   Tracking$^b$ (extended, 0.23$^\circ$) & +1.5   & $<1.28\times 10^{-11}$ \\
   EGRET$^c$ (point)                     & +2.8   & $<1.14\times 10^{-11}$\\
   EGRET$^d$ (extended, 0.5$^\circ$)     & +1.4   & $<1.95\times 10^{-11}$\\ 
   NML$^e$ (point)                       & $-$0.7 &$<4.47\times 10^{-12}$\\ \hline \hline
   \multicolumn{3}{l}{\scriptsize a: Radio VLBI core position \cite{Johnston:1}. RA (J2000) 13$^h$25$^m$29$^s$ Dec -43$^\circ$01$^m$12$^s$} \\
   \multicolumn{3}{l}{\scriptsize b: Radius 14$^\prime$(0.23$^\circ$) covering ROSAT PSPC emission \cite{Turner:1}.}\\
   \multicolumn{3}{l}{\scriptsize c:  Highest significance within the error circle (0.68$^\circ$ \cite{Nolan:1}). RA 13$^h$23$^m$15$^s$ Dec -43$^\circ$31$^m$12$^s$.}\\
   \multicolumn{3}{l}{\scriptsize d: Source radius limited by $\pm 1^\circ$ search. RA 13$^h$26$^m$02$^s$ Dec -43$^\circ$31$^m$12$^s$. }\\
   \multicolumn{3}{l}{\scriptsize e:  Northern Middle Lobe. Point source at max radio position \cite{Junkes:1}. RA 13$^h26^m08^s$ Dec -42$^\circ15^m35^s$ }\\  
  \end{tabular}
  \end{center}
 \end{table}
 
\section{Results and Discussion}
 No significant excesses were found within our search region and 3$\sigma$ upper limits were 
 derived (table~\ref{tab:upplimits}) for the features discussed earlier.
 The upper limits from this work are not in conflict with previous measurements, and 
 lie at least an order of magnitude above 
 the extrapolated EGRET flux (integral spectral index $-$1.5, \cite{Nolan:1}) at 1.5 TeV.
 We therefore cannot place constraints on gamma ray emission models concerning Centaurus~A with
 this dataset.  
 The upper limit from the point source within the EGRET error circle must be considered with a statistical
 penalty of $\sim$100 due to the non {\em a priori} nature of the search. 
 We note that our observations 
 were likely taken during a low state of X-ray emission, and might expect a greater chance of
 detectable TeV emission during a high state. The detection of TeV gamma rays by \cite{Grindlay:1} was
 achieved during the historically highest state of X-ray emission. Clearly, observations are required with 
 more sensitive telescopes at energies below 1 TeV, for example, by CANGAROO II \cite{Yoshikoshi:1} and HESS 
 \cite{Aharonian:1}, covering all states of emission.

\begin{ack}
 This work is supported by a Grant-in-Aid in Scientific Research from the Japanese Ministry
 of Science, Sports and Culture, and also by the Australian Research Council. GR, MR \& TY acknowledge
 the receipt of JSPS postdoctoral fellowships.
\end{ack}

\end{document}